\documentclass[9pt, conference]{IEEEtran}
\IEEEoverridecommandlockouts
\usepackage{cite}
\usepackage{amsmath,amssymb,amsfonts}
\usepackage{algorithmic}
\usepackage[algo2e,ruled,vlined,linesnumbered]{algorithm2e}
\usepackage{tabularx}
\usepackage{adjustbox}
\usepackage{graphicx}
\usepackage{textcomp}
\usepackage{xcolor}
\usepackage{booktabs}
\usepackage[nobiblatex]{xurl}
\usepackage{hyperref}
\usepackage{cleveref}
\usepackage{enumitem}

\usepackage{tikz}

\newcommand*\circled[1]{\tikz[baseline=(char.base)]{
            \node[shape=circle,draw,inner sep=0.8pt, minimum size=2pt] (char) {#1};}}

\newcommand{\rpoint}[1]{\circled{{\fontfamily{pcr}\selectfont\footnotesize{#1}}}}

\def\BibTeX{{\rm B\kern-.05em{\sc i\kern-.025em b}\kern-.08em
    T\kern-.1667em\lower.7ex\hbox{E}\kern-.125emX}}

\usepackage[compact]{titlesec}

\titlespacing\section{0pt}{0.3\baselineskip}{0.2\baselineskip}
\titlespacing\subsection{0pt}{0.2\baselineskip}{0.1\baselineskip}
\titlespacing\subsubsection{0pt}{0.15\baselineskip}{0.1\baselineskip}


\usepackage{fancyhdr,lipsum}
\fancypagestyle{firstpage}{% Page style for first page
  \fancyhf{}% Clear header/footer
  \fancyhead[C]{To appear at the IEEE International Conference on Quantum Artificial Intelligence (QAI), Naples, Italy, November 2025.}% Header
  \fancyfoot[C]{\thepage}% Footer
}

\begin{document}

\title{Cutting is All You Need: Execution of Large-Scale Quantum Neural Networks on Limited-Qubit Devices
}

\author{%\small
\IEEEauthorblockN{Alberto Marchisio\textsuperscript{1,2}, Emman Sychiuco\textsuperscript{1,2}, Muhammad Kashif\textsuperscript{1,2}, Muhammad Shafique\textsuperscript{1,2}\\}
\IEEEauthorblockA{\textit{\textsuperscript{1}eBRAIN Lab, Division of Engineering} \textit{New York University (NYU) Abu Dhabi}, Abu Dhabi, UAE\\
\textit{\textsuperscript{2}Center for Quantum and Topological Systems (CQTS), NYUAD Research Institute, New York University Abu Dhabi}, UAE\\
Emails: \{alberto.marchisio, jes9843, muhammadkashif, muhammad.shafique\}@nyu.edu 
}}

%Alberto, Emman, Kashif, Shafique

%Track: A5 Applications of Emerging Technologies 

% \and
% \and
% \and
% \and
% \and
\thispagestyle{empty}
\maketitle
\thispagestyle{empty}
\pagenumbering{gobble}
\thispagestyle{firstpage}

\begin{abstract}
The rapid advancement in Quantum Computing, particularly through Noisy-Intermediate Scale Quantum (NISQ) devices, has spurred significant interest in Quantum Machine Learning (QML) applications. Despite their potential, fully-quantum algorithms remain impractical due to the limitations of current NISQ devices. Hybrid quantum-classical neural networks (HQNNs) have emerged as a viable alternative, leveraging both quantum and classical computations to enhance machine learning capabilities. However, the constrained resources of NISQ devices, particularly the limited number of qubits, pose significant challenges for executing large-scale quantum circuits.

This work addresses these current challenges by proposing a novel and practical methodology for quantum circuit cutting of HQNNs, allowing large quantum circuits to be executed on limited-qubit NISQ devices. Our approach not only preserves the accuracy of the original circuits but also supports the training of quantum parameters across all subcircuits, which is crucial for the learning process in HQNNs. We propose a cutting methodology for HQNNs that employs a greedy algorithm for identifying efficient cutting points, and the implementation of trainable subcircuits, all designed to maximize the utility of NISQ devices in HQNNs. The findings suggest that quantum circuit cutting is a promising technique for advancing QML on current quantum hardware, since the cut circuit achieves comparable accuracy and much lower qubit requirements than the original circuit. The code is available at \href{https://github.com/eBrain4Everyone/QNN-Cutting}{https://github.com/eBrain4Everyone/QNN-Cutting}.
\end{abstract}

\begin{IEEEkeywords}
Quantum Neural Networks, Circuit Cutting, Quantum Machine Learning
\end{IEEEkeywords}

\section{Introduction}

Quantum computing has demonstrated significant advancements, particularly with Noisy-Intermediate Scale Quantum (NISQ) devices~\cite{Arute_2019Nature_QuantumSupremacy, Araujo_2020SR_DivideAndConquer, Huang_2022Science_QuantumAdvantageExperiments, Pokharel_2023PRL_QuantumSpeedup, Kim_2023Nature_UtilityQuantumIBM, Tudisco_2024Access_AdvantagesQuantumFraudDetection, kashif2022demonstrating, innan2025next}. However, due to the constraints of NISQ hardware, such as limited qubits, noise, and scalability issues, fully Quantum Machine Learning (QML) algorithms remain impractical~\cite{Preskill_2018arxiv_QC_NISQ_era,kashif2021design}. Hybrid Quantum-Classical Neural Networks (HQNNs) have emerged as a promising alternative, combining classical and quantum layers for machine learning tasks~\cite{Zaman_2024arxiv_QMLSurvey, Chen_2022PRR_QCNN, Masum_2023ICMLA_HQNN_SentimentAnalysis, LAbbate_2024TQE_QCTraining}. HQNNs leverage quantum circuits for complex representations while retaining classical robustness, making them suitable for NISQ-era hardware~\cite{Hu_2022ICCD_QGCNN_NISQ, Kashif_2024arxiv_InvestigatingNoiseHQNNs}.

%and employ quantum circuits that interface between classical and quantum computations~\cite{Bergholm_2018arxiv_PennyLane}. Angle embedding and amplitude embedding are commonly used to map classical data into quantum states~\cite{Schuld_2018Springer_SupervisedQML, Rath_2023arxiv_QuantumDataEncoding}. The angle embedding method consists of rotating each classical feature by a certain angle based on its magnitude, while the amplitude embedding encodes classical data into the amplitudes of a quantum state. The angle embedding, due to its 1-to-1 representation with the number of features into qubit states, is known for its high precision. On the other hand, using the amplitude encoding, $n$ qubits are sufficient to encode $2^n$ features~\cite{Schuld_2018Springer_SupervisedQML}. Hence, it can encode the information in a compact form.

\subsection{Target Research Problem and Associated Challenges}
\label{subsec:Intro_TargetProblems}

Despite their potential, HQNNs face scalability challenges due to qubit limitations. Large-scale quantum circuits suffer from noise\cite{kashif:2025_HQNET,kashif2024nrqnn,ahmed2025noisyhqnns,ahmed2025quantum}, barren plateaus~\cite{McClean_2018arxiv_BarrenPlateaus,kashif2023impact,kashif2024resqnets,kashif2024alleviating,kashif2025deep}, and significant computational overhead. Quantum circuit cutting~\cite{Peng_2020PRL_LargeQuantumCircuitsSmallQuantumCOmputer} is a promising technique that partitions large circuits into smaller subcircuits, enabling execution on limited-qubit devices.

However, while circuit cutting has shown success in other quantum computing applications, applying it to trainable models like HQNNs introduces a fundamentally different set of challenges. Existing cutting techniques~\cite{Tang_2021ASPLOS_CutQC, Lowe_2023Quantum_FastQuantumCircuitCutting, Piveteau_2024TIT_CircuitKnitting, Gentinetta_2024Quantum_CircuitKnitting, Bechtold_2023QST_EffectCircuitCuttingQAOANISQ, Brandhofer_2024TQE_PartitioningQuantumCircuitsGateCutsWireCuts, Schmitt_2024arxiv_CuttingCircuitsTwoQubitUnitaries, Harrow_2024arxiv_OptimalQuantumCircuitCuts, Bechtold_2024IPDPSW_CuttingWire, Chen_2024TQC_QuantumCircuitCuttingClassicalShadows, Kan_2024arxiv_ScalableCircuitCutting, Sarker_2024IPDPSW_PerformanceWireCircuitCuttingEntanglements, Pednault_2023arxiv_AlternativeWireCuttingAncillaQubits, Harada_2023arxiv_DoublyOptimalParallelWireCutting}
are incompatible with gradient-based training because they lack support for end-to-end differentiability. This limitation is critical: training HQNNs requires propagating gradients through quantum subcircuits, a feature incompatible with traditional cutting approaches. For example, current automatic differentiation frameworks like PennyLane~\cite{Bergholm_2018arxiv_PennyLane} are limited to differentiating a single quantum node (QNode) by converting it into a differentiable Torch or Keras layer. However, a cut circuit in HQNNs spans multiple QNodes, breaking the gradient flow required for backpropagation.

This gap motivates the need for a separate and focused investigation into circuit cutting tailored for HQNNs. Unlike conventional circuit cutting, which optimizes for execution or fidelity, cutting in the context of HQNNs must preserve differentiability, model integrity, and training dynamics. As such, circuit cutting for HQNNs is not merely an extension of prior techniques, but it is a distinct problem that necessitates new strategies for partitioning and gradient management.

\textbf{Summary of Challenges:}

\begin{itemize}[leftmargin=*]
    \item \textit{Limited qubit availability:} Large-scale quantum circuits require more qubits than available in current NISQ devices.
    \item \textit{Gradient propagation in cut circuits:} Existing quantum circuit cutting techniques do not support backpropagation across subcircuits, preventing effective training of HQNNs.
    \item \textit{Computational efficiency:} While circuit cutting enables execution on limited-qubit devices, it introduces overhead, requiring an optimized approach for balancing accuracy and efficiency.
\end{itemize}

To overcome these challenges, we propose a framework that (i) enables quantum circuit cutting for HQNNs, (ii) maintains trainability by supporting gradient backpropagation across subcircuits, and (iii) efficiently partitions circuits under hardware constraints.

\subsection{Motivational Case Study}
\label{subsec:Motivational_Analysis}

\begin{figure}[!t]
  \centering
  \includegraphics[width=\linewidth]{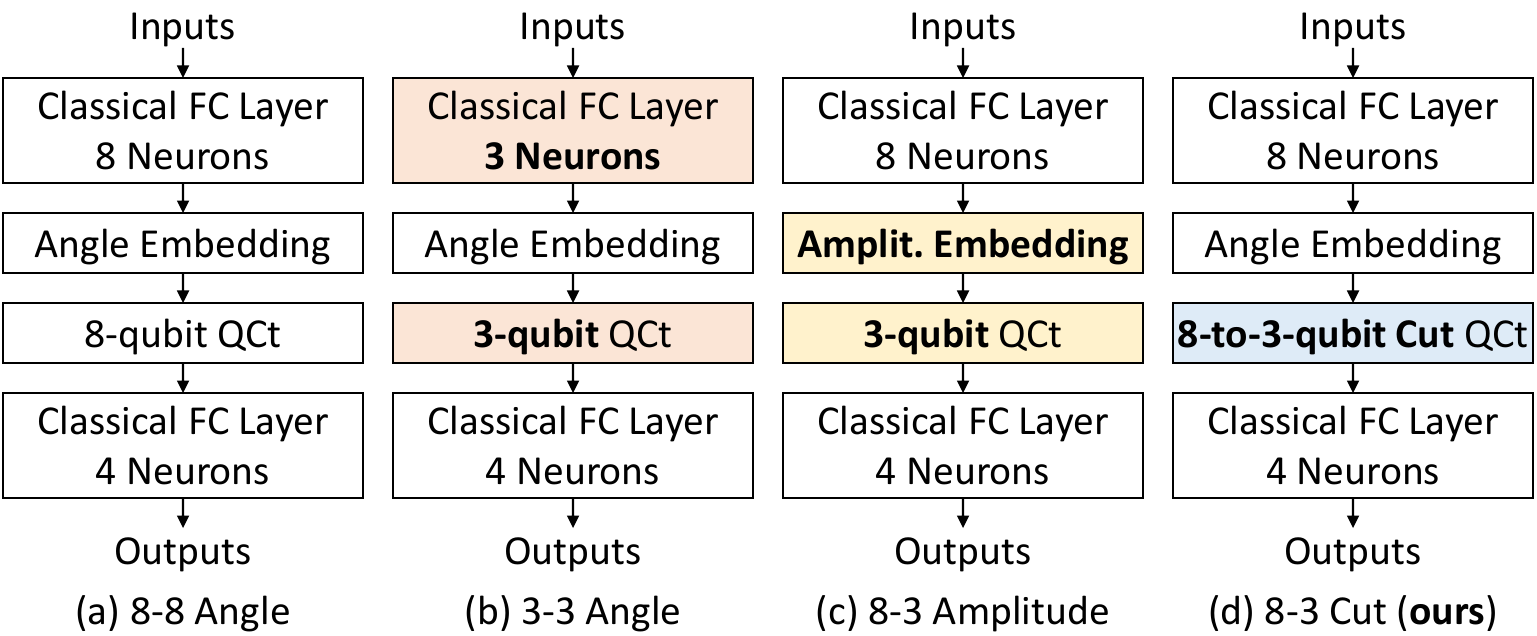}
  \caption{Variants of the HQNN for this experiment. (a) Original HQNN, with an 8-neuron FC layer, angle embedding, 8-qubit quantum circuit (QCt), and 4-neuron FC layer. (b) Modified HQNN with reduced features, which has a 3-neuron FC layer and 3-qubit QCt. (c) Modified HQNN with amplitude embedding and 3-qubit QCt. (d) Modified HQNN with 8-to-3-qubit cut QCt.}
  \label{fig:motiv_analysis_techniques}
\end{figure}

We evaluate four HQNN variants on the Digits dataset~\cite{Alpaydin_1998UCI_DigitsDataset} (Fig.~\ref{fig:motiv_analysis_techniques}). The original HQNN (\textit{8-8 Angle}) contains an 8-neuron fully connected (FC) layer, followed by an 8-qubit quantum circuit (QCt), and a 4-neuron FC layer. The 8 classical features at the output of the first classical FC layer are encoded into quantum states with angle embedding\footnote{Angle embedding and amplitude embedding are commonly used to map classical data into quantum states~\cite{Schuld_2018Springer_SupervisedQML, Rath_2023arxiv_QuantumDataEncoding}. The angle embedding method consists of rotating each classical feature by a certain angle based on its magnitude, while the amplitude embedding encodes classical data into the amplitudes of a quantum state. The angle embedding, due to its 1-to-1 representation with the number of features into qubit states, is known for its high precision. On the other hand, using the amplitude encoding, $n$ qubits are sufficient to encode $2^n$ features~\cite{Schuld_2018Springer_SupervisedQML}. Hence, it can encode the information in a compact form.}. Given an available \mbox{3-qubit} quantum device, we explore:
\begin{itemize}[leftmargin=*]
    \item \textit{3-3 Angle}: Reducing the first FC layer to 3 neurons, maintaining angle embedding.
    \item \textit{8-3 Amplitude}: Encoding 8 features into 3 qubits using amplitude embedding.
    \item \textit{8-3 Cut}: Applying quantum circuit cutting to execute an 8-qubit circuit on a 3-qubit device.
\end{itemize}

\begin{figure}[!t]
  \centering
  \includegraphics[width=\linewidth]{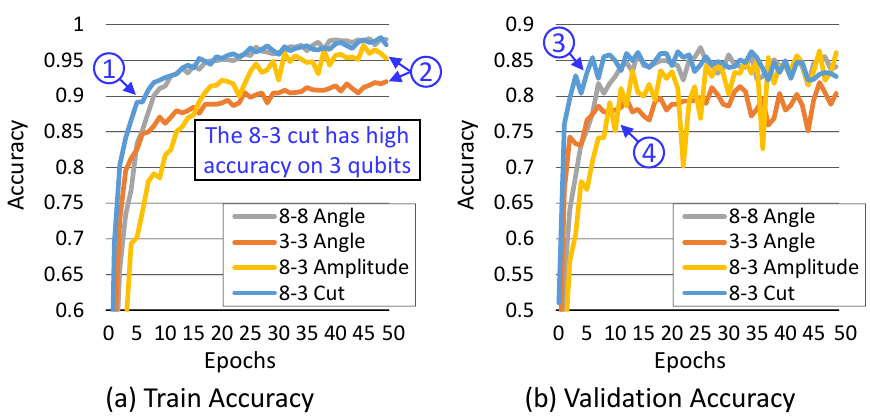}
  \caption{Training and validation accuracy for different HQNN architectures.}
  \label{fig:motiv_analysis}
\end{figure}

The experimental results shown in \Cref{fig:motiv_analysis} demonstrate that the 8-3 Cut architecture preserves accuracy while reducing qubit requirements. The 3-3 Angle and 8-3 Amplitude architectures suffer from feature loss and encoding inefficiencies, leading to lower accuracy (see label~\rpoint{2}). As indicated by labels~\rpoint{1} and~\rpoint{3}, the 8-3 Cut HQNN converges faster than the original 8-8 Angle HQNN in early training epochs. The lower accuracy of the 3-3 Angle HQNN stems from information loss, as only 3 features are propagated instead of 8. The 8-3 Amplitude HQNN underperforms compared to the 8-8 Angle and 8-3 Cut architectures due to the different embedding strategy. Moreover, as highlighted by label~\rpoint{4}, it requires more epochs to achieve high validation accuracy.

This analysis highlights that \textit{quantum circuit cutting enables the execution of quantum circuits on limited-qubit devices while maintaining accuracy comparable to the original HQNN architecture}.

%This simple yet insightful analysis demonstrates that \textit{the approach with quantum circuit cutting has the potential to execute quantum circuits on limited-qubit devices with minimal changes to the rest of the HQNN architecture, maintaining similar accuracy levels compared to the original circuit}.

\subsection{Novel Contributions}

To solve the above-discussed challenges, we make the following novel contributions (see \Cref{fig:Novel_contrib}):

\begin{figure}[!t]
  \centering
  \includegraphics[width=\linewidth]{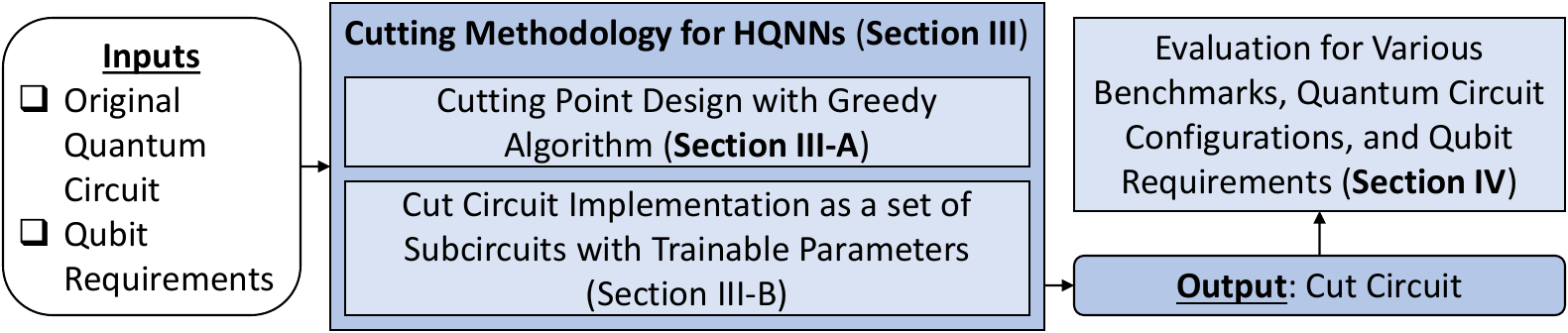}
  \caption{Overview of the novel contributions in this work.}
  \label{fig:Novel_contrib}
\end{figure}

\begin{itemize}[leftmargin=*]
    \item We propose a framework for cutting HQNNs into multiple subcircuits, to enable their full execution on limited-qubit devices. (\textbf{\Cref{sec:methodology}})
    \item We design a greedy algorithm to identify the cutting points based on the given qubit constraints. %to automatically identify the cutting points in a given quantum circuit under a given qubit constraint for the underlying quantum device. 
    (\textbf{\Cref{subsec:methodology_cutting_points_design}})
    \item We implement the cut circuit as a set of subcircuits in a way that each node can be trained during the HQNN learning process. (\textbf{\Cref{subsec:methodology_cut_circuit_implementation}})
    \item We evaluate our methodology across various benchmarks, demonstrating accuracy preservation with reduced qubit requirements. %and configurations of quantum circuits and qubit requirements. The experimental results demonstrate that the cut circuits achieve comparable accuracy to the original circuit with a small computation overhead, thereby enabling the execution of more complex circuits on limited-qubit devices. 
    (\textbf{\Cref{sec:experiments}})
    \item For reproducible research, we release a repository containing the code of the complete framework with cut circuits at \href{https://github.com/eBrain4Everyone/QNN-Cutting}{https://github.com/eBrain4Everyone/QNN-Cutting}.
\end{itemize}

By enabling large-scale HQNNs on constrained quantum hardware, our approach paves the way for practical QML in the NISQ era.

\section{Background and Related Work}
\label{sec:Background}

\subsection{Hybrid Quantum-Classical Neural Networks}

HQNNs are hybrid systems that combine classical and quantum computational elements, making them well-suited for NISQ devices~\cite{Arthur_2022QCE_HQNNs, Bhowmik_2024arxiv_TransferLearningHQNN, Wang_2024ASPDAC_JustQ}. In an HQNN, classical and quantum layers are integrated, typically in a sequential manner, allowing for a synergistic use of both types of computational resources~\cite{Hafeez_2024AI_HQNN, zaman2024comparative, kashif2024computational}. Classical layers handle preprocessing, feature extraction, and output, while quantum layers perform complex operations like quantum feature mapping and entanglement. Similarly to the weights of the classical layers, the gates of the parameterized quantum circuits (PQCs) contain trainable parameters~\cite{Senokosov_2024MLST_QMLImageClassification}. 

Classical data is transitioned into quantum layers by encoding it into quantum states through techniques such as angle embedding, which maps features to qubit rotations, or amplitude embedding, which compresses data into the quantum state’s amplitudes~\cite{Schuld_2018Springer_SupervisedQML}. The quantum layers then process this data, leveraging high-dimensional feature spaces to analyze complex patterns. After quantum operations, the results are measured and fed back to the classical layers for further processing or decision-making~\cite{Zaman_2024arxiv_StudyingImpactQuantumHyperparameters}. 
The hybrid structure of HQNNs takes advantage of quantum computing's strengths, such as parallelism and entanglement, while the classical layers mitigate the limitations of current quantum hardware~\cite{Schillo_2024arxiv_QuantumCircuitLearningNISQ, el2025designing}. This design enhances robustness against noise and errors and allows for greater scalability, making HQNNs a promising approach for advancing quantum machine learning, especially when fully-quantum networks are not yet feasible~\cite{Arthur_2022QCE_HQNNs}.

\begin{figure}[!t]
  \centering
  \includegraphics[width=\linewidth]{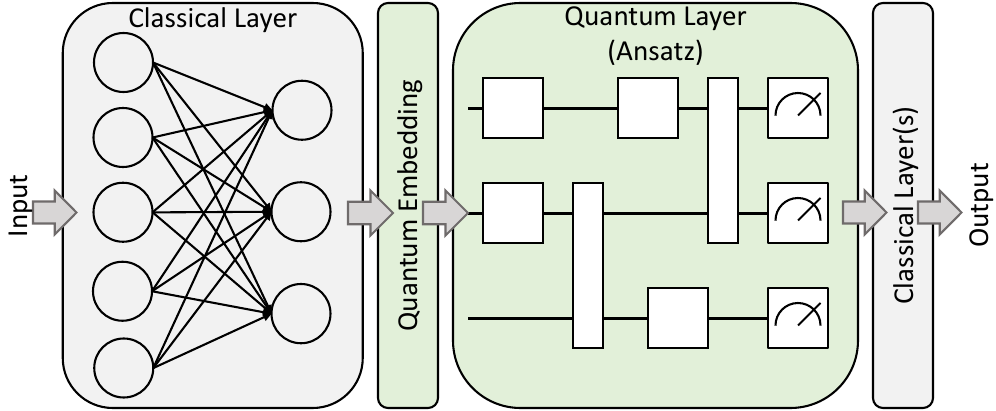}
  \caption{Example of HQNN architecture, where a quantum layer is interleaved between classical layers.}
  \label{fig:HQNN_model}
\end{figure}

\Cref{fig:HQNN_model} illustrates an example of an HQNN, where the input is processed through a classical layer, followed by the quantum embedding to encode the classical features into quantum states. After that, the quantum circuit (Ansatz) contains a set of gates that modify the state of the qubits, and the measurement operations read the values to be passed to the following classical layers. Finally, a set of one or more classical layers generates the output predictions. The outputs of the HQNNs are plugged into a loss function to train the parameters through classical backpropagation~\cite{Beer_2020Nature_TrainingDeepQNNs}.

\subsection{Quantum Circuit Cutting}

\begin{figure}[!ht]
  \centering
  \includegraphics[width=\linewidth]{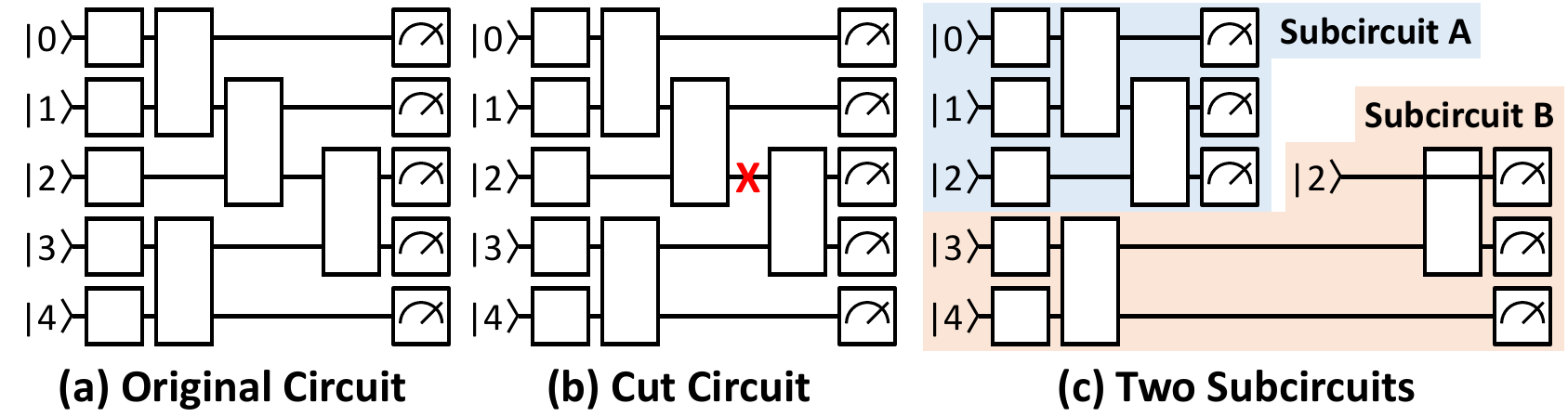}
  \caption{Example of cutting a 5-qubit circuit to fit in a 3-qubit device. (a) Original circuit. (b) Cut circuit, where the cut is represented by a red cross. (c) Representation of the cut circuit as two 3-qubit subcircuits.}
  \label{fig:cutting_background}
\end{figure}

Quantum circuit cutting has emerged as a solution to execute large circuits on small devices by breaking quantum wires using a measurement operation to save the values classically and reconstructing them across subcircuits by re-encoding the values into quantum states~\cite{Peng_2020PRL_LargeQuantumCircuitsSmallQuantumCOmputer}. \Cref{fig:cutting_background} shows an example of this mechanism for a 5-qubit circuit implemented on 3-qubits.

%Motivated by the need to execute large quantum circuits on small quantum devices, the quantum circuit cutting technique has been developed~\cite{Peng_2020PRL_LargeQuantumCircuitsSmallQuantumCOmputer}. The key idea is to break the quantum wires using a measurement operation to save the value classically and embedding operation to encode the value into a quantum state. \Cref{fig:cutting_background} shows an example of this mechanism for a 5-qubit circuit implemented on 3-qubits.

In recent years, several variants and optimizations of the quantum circuit cutting method have been proposed in the literature~\cite{Tang_2021ASPLOS_CutQC, Lowe_2023Quantum_FastQuantumCircuitCutting, Piveteau_2024TIT_CircuitKnitting, Gentinetta_2024Quantum_CircuitKnitting, Bechtold_2023QST_EffectCircuitCuttingQAOANISQ, Brandhofer_2024TQE_PartitioningQuantumCircuitsGateCutsWireCuts, Schmitt_2024arxiv_CuttingCircuitsTwoQubitUnitaries, Harrow_2024arxiv_OptimalQuantumCircuitCuts, Bechtold_2024IPDPSW_CuttingWire, Chen_2024TQC_QuantumCircuitCuttingClassicalShadows, Kan_2024arxiv_ScalableCircuitCutting, Sarker_2024IPDPSW_PerformanceWireCircuitCuttingEntanglements, Pednault_2023arxiv_AlternativeWireCuttingAncillaQubits, Harada_2023arxiv_DoublyOptimalParallelWireCutting}. Moreover, quantum circuit cutting has been used for error mitigation~\cite{Li_2023ICCD_VirtualDistillationCircuitCuttingMitigation, Majumdar_2022arxiv_ErrorMitigatedQuantumCircuitCutting, Basu_2024JSS_FragQC} and intellectual property protection~\cite{typaldos2024leveraging}.

However, these works are applied to quantum algorithms but not to QML and HQNNs. This is a major limitation because the current PennyLane implementations that convert quantum circuits to Keras layers~\footnote{\url{https://pennylane.ai/qml/demos/tutorial_qnn_module_tf/}} or Torch layers\footnote{\url{https://pennylane.ai/qml/demos/tutorial_qnn_module_torch/}} require as input a single QNode. This is a fundamental step in HQNNs that allows to have trainable parameters in the quantum circuit. Since the cut circuit cannot be implemented on a single QNode but requires multiple QNodes (one for each subcircuit), the quantum circuit cutting in its form cannot be adopted in HQNNs. \textit{To overcome this issue, our framework builds a QNode for each subcircuit and converts each QNode to a trainable Keras layer.} 

On the other hand, the works in~\cite{Marshall_2023Quantum_HighDimensionalQMLSmallQuantumComputers, Bilek_2024arxiv_UtilizingSmallQuantumComputersML} are applied to QML but they only focus on quantum circuit partitioning. The works in~\cite{pira2022invitationdistributedquantumneural, Kawase_2024QMI_DistributedQNNs, Innan_2024arxiv_FedQNN} discuss how to execute HQNNs on distributed systems. However, the scope of our work is to provide a platform for employing quantum circuit cutting on HQNNs.

%FCM: A Fusion-aware Wire Cutting Approach for Measurement-based Quantum Computing

\section{HQNN Quantum Circuit Cutting Methodology}
\label{sec:methodology}

%\subsection{Overview}
%\label{subsec:methodology_overview}

\begin{figure*}[!ht]
  \centering
  \includegraphics[width=.9\linewidth]{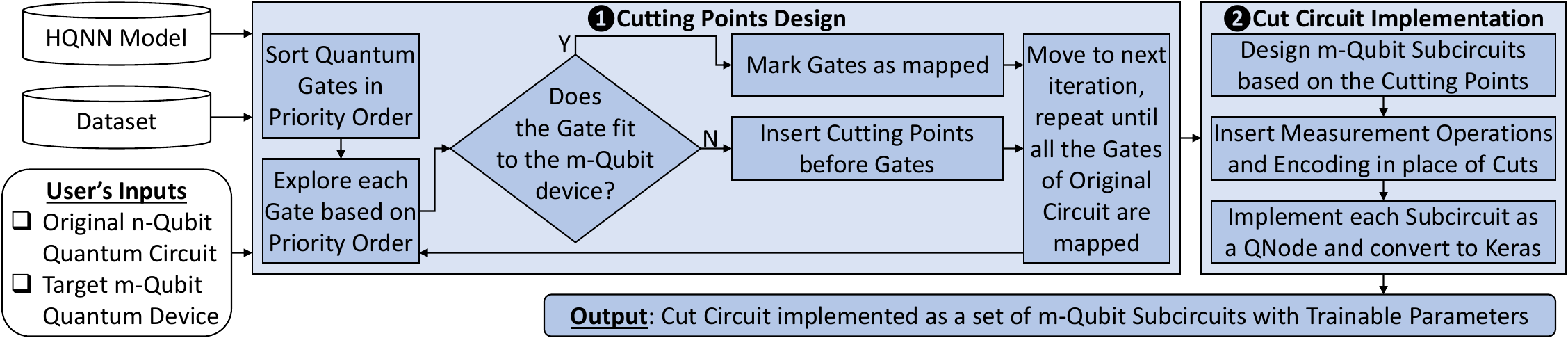}
  \caption{Our methodology for cutting point design and implementation for HQNNs.}
  \label{fig:methodology_overview}
\end{figure*}

Our methodology, depicted in \Cref{fig:methodology_overview}, starts with the original n-qubit quantum circuit and targets an m-qubit quantum device. The key steps involve designing optimal cutting points, generating subcircuits, and ensuring trainable parameters across the entire HQNN.

%Our methodology for designing and implementing quantum circuit cutting is depicted in \Cref{fig:methodology_overview}. The inputs to the framework are the dataset, the HQNN model, which contains the information on the original n-qubit quantum circuit, and the target m-qubit quantum device. 

%The first step of the process, which is described in \Cref{subsec:methodology_cutting_points_design}, consists of designing the cutting points in the quantum circuit. After sorting the quantum gates in priority order, for each iteration, the cutting points are inserted before gates that have dependencies with other states that are not available and/or do not fit in the current m-qubit device but will be covered in one of the following iterations of the algorithm, i.e., will be mapped to another subcircuit. Afterward, \Cref{subsec:methodology_cut_circuit_implementation} describes how the implementation of the cut circuit is conducted. The subcircuits are generated based on the cut points to form the cut circuit. Every cut point is replaced with a measurement operation in the preceding subcircuit and an encoding operation in the following subcircuit. Each subcircuit is implemented as an independent QNode, in such a way that the complete cut circuit with trainable parameters can be trained through the traditional backpropagation process.

\subsection{Cutting Points Design}
\label{subsec:methodology_cutting_points_design}

%Describe it in the form of a greedy algorithm that selects the cutting points. (Algorithm + text)

%We implement a greedy algorithm to identify cutting points based on circuit dependencies and qubit availability (\Cref{alg:CuttingPointsDesign}). Gates are mapped to subcircuits with priority given to data dependencies. Gates with unresolved dependencies trigger cuts, ensuring all subcircuits fit within the target device’s qubit capacity.

The \textit{Cutting Points Design Algorithm}, detailed in \Cref{alg:CuttingPointsDesign}, aims to identify optimal cutting points within a quantum circuit, enabling its execution on devices with a lower number of qubits than the original circuit. The algorithm starts by sorting the quantum gates based on their priority, determined by data dependencies within the circuit. Initially, the set of gates is divided into two groups: \texttt{UnmappedGates}, which contains all the gates to be processed, and \texttt{MappedGates}, which will store gates that have been successfully assigned to subcircuits.

The algorithm iterates by selecting \( n \) wires at a time, forming a subcircuit based on the device’s qubit limit. For each gate in the sorted list, the algorithm checks whether it can be implemented in the selected \( n \)-qubit subcircuit, while ensuring all its dependencies are satisfied. If all dependencies are available, the gate is moved from \texttt{UnmappedGates} to \texttt{MappedGates}. However, if a gate can be executed on the \( n \)-qubit subcircuit but has unresolved dependencies, the algorithm places a cut before this gate and proceeds to avoid mapping subsequent gates on the same wire within the current subcircuit. This ensures the circuit remains properly segmented based on qubit availability. 
The process repeats until all gates are successfully mapped, yielding a series of cut subcircuits, each capable of execution on a \( n \)-qubit device. %This approach optimizes circuit execution for NISQ-era devices, balancing the number of cuts with computational feasibility.

\begin{algorithm2e}[!ht]
\SetAlgoLined
\begin{small}
\SetKwFunction{CuttingPointsDesign}{CuttingPointsDesign}
\SetKwProg{procCPD}{Procedure}{}{end}
\procCPD{\CuttingPointsDesign{Gates, n}}
{
    Sort Gates in Priority Order, from $1$ to $MaxGate$; \tcp*[h]{priority based on the data dependencies in the circuit}\\
    $UnmappedGates \leftarrow Gates$;\\
    $MappedGates \leftarrow \emptyset$;\\
    \While{$UnmappedGates \neq \emptyset$}
    {
        Select $m$ wires based on priority; \tcp*[h]{This would form a m-qubit subcircuit}\\
        \For{$UnmappedGates \in \{1, ... MaxGate\}$}
        {
            \uIf{$UnmappedGates_i$ can be implemented in the $n$ wires and all the dependencies are available}
            {
                Move $UnmappedGates_i$  to $MappedGates_i$;
            }
            \uElseIf{$UnmappedGates_i$ can be implemented in the $n$ wires but at least one dependency is not available}
            {
                Place a cut before $UnmappedGates_i$;\\
                Do not consider the following gates on this wire as implementable in the $n$ wires;
            }
        }
    }
}
\end{small}
\caption{Cutting Points Design for Quantum Circuits}
\label{alg:CuttingPointsDesign}
\end{algorithm2e}

\subsection{Cut Circuit Implementation}
\label{subsec:methodology_cut_circuit_implementation}

%Subcircuits are generated by replacing cuts with measurement operations and re-encoding quantum states. Each subcircuit becomes an independent QNode integrated into the HQNN as quantum layers. The full HQNN is trained using backpropagation, where trainable parameters are adjusted across all subcircuits, enabling gradient-based learning.

\begin{figure}[!ht]
  \centering
  %\vspace*{-5pt}
  \includegraphics[width=\linewidth]{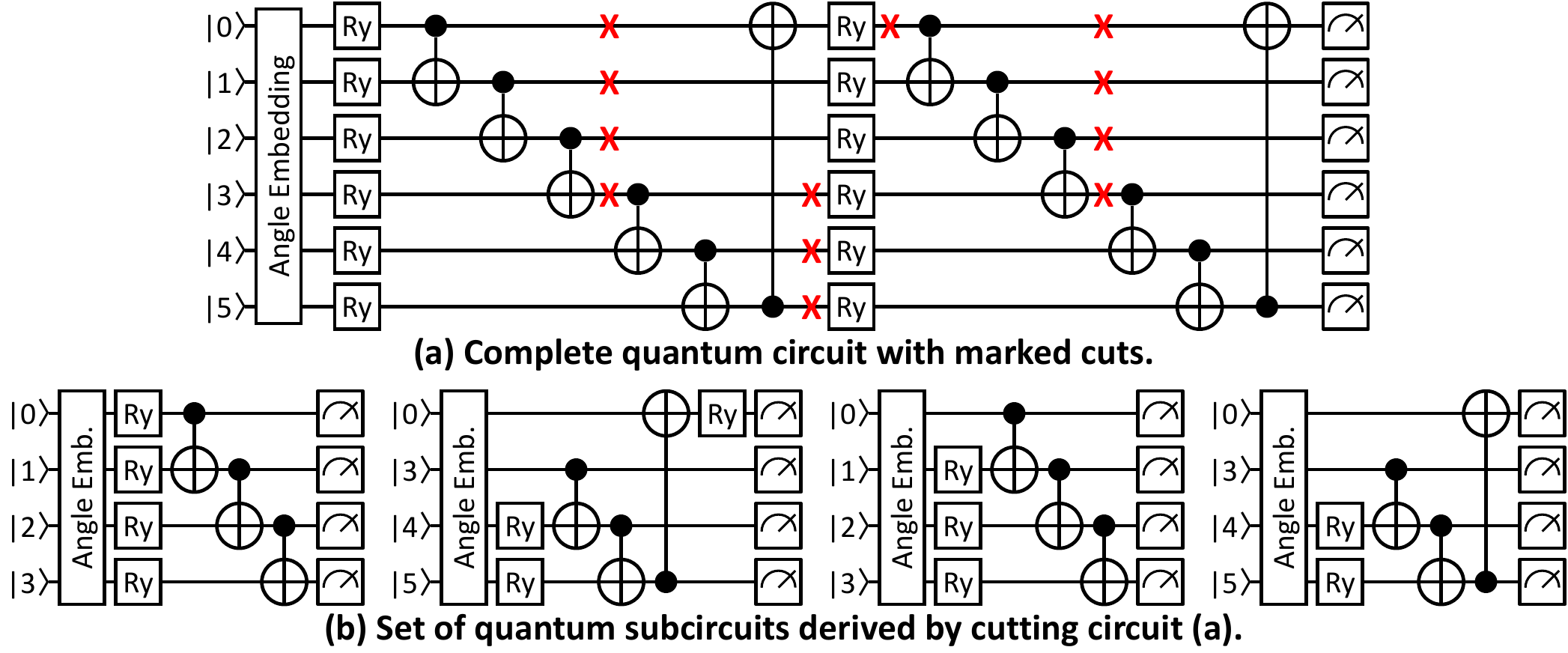}
  \caption{Example of the generation of subcircuits in a 6-4 cut, given the original circuit and the cut placements.}
  \label{fig:cutting_example_6to4}
\end{figure}

Our implementation of the cut circuit demonstrates the process of cutting the quantum circuit and integrating its subcircuits in the correspondent HQNN, having defined the cut placements within the original circuit. Subcircuits are generated by replacing cuts with measurement operations and re-encoding quantum states (using angle embedding). An example of subcircuit generation, based on the cuts in the original circuit, is shown in \Cref{fig:cutting_example_6to4}.

%, the set of subcircuits is formed to ensure that each subcircuit can be executed within the qubit limits of the hardware while preserving the structure of the original quantum circuit. Each cut is replaced with a measurement operation and an encoding operation (using angle embedding) in the following subcircuit. An example of subcircuit generation, based on the cuts in the original circuit, is shown in \Cref{fig:cutting_example_6to4}.

After constructing the quantum circuit and applying the necessary cuts, the full HQNN model is assembled. The classical layers, are interwoven with the quantum subcircuits. Each subcircuit is described as an independent QNode and properly connected to each other to form the complete HQNN. In this way, all the subcircuits of the HQNN can be integrated into the network as quantum layers using PennyLane's \texttt{KerasLayer}, which manages the parameterized quantum gates and their respective operations, using the traditional backpropagation mechanism. 

During training, both the forward and backward passes are profiled to compute the number of floating-point operations (FLOPs), allowing for an analysis of the computational cost of the cut circuit implementation. Moreover, the training and validation accuracy values are measured and recorded for every training epoch.

\section{Evaluation of our Cutting Methodology}
\label{sec:experiments}

\subsection{Experimental Setup}
\label{subsec:exp_setup}

%Experiments are conducted on the Digits~\cite{Alpaydin_1998UCI_DigitsDataset} and MNIST~\cite{Deng2012MNIST} datasets. The quantum circuits are implemented using PennyLane~\cite{Bergholm_2018arxiv_PennyLane}, and training is performed on an Nvidia RTX 6000 Ada GPU. HQNN architecture consists of one classical layer, a quantum layer with two repetitions of the entangling Ansatz, and another classical layer. All experiments are repeated five times to mitigate randomness.

The experiments are conducted on the Digits~\cite{Alpaydin_1998UCI_DigitsDataset} and MNIST~\cite{Deng2012MNIST} datasets. Note that this is a typical settings for benchmarking HQNNs. The quantum circuits are implemented in PennyLane~\cite{Bergholm_2018arxiv_PennyLane} with the Tensorflow Keras backend. The experiments run on a system equipped with an NVIDIA GeForce RTX 3060 GPU and AMD Ryzen 7 5800H CPU. The HQNN architecture consists of one classical fully-connected layer (where the number of neurons is equal to the number of qubits of the original quantum circuit), followed by a quantum layer with two repetitions of the basic entangling Ansatz, and another classical fully-connected layer with ten neurons (to match the number of classes of the datasets). The cut circuit experiments have been conducted sequentially on the same quantum system, which is reset and initialized iteratively for each subcircuit. All experiments are repeated five times to mitigate randomness, 
%To mitigate the effects of randomness in initialization seeds, each training run is repeated five times, 
and the results show the average accuracy over these five runs. A detailed list of the experiment setup details and hyperparameters is visualized in \Cref{tab:exp_setup}.

\begin{table}[!ht]
\centering
\caption{Experiment Setup Details}
\label{tab:exp_setup}
\begin{adjustbox}{max width=.8\linewidth}
\begin{tabular}{c|c}
\textbf{Item} & \textbf{Experiment Detail} \\ \toprule
Quantum Computing FrameWork & \verb|PennyLane| \\ \midrule
%Back-End Simulator & \verb| lightning.qubit(PL)|, \verb|qasm_simulator(QK)|  \\ \midrule
Back-End GPU Machine & \verb| NVIDIA GeForce RTX 3060 |\\ \midrule
Deep Learning Interface & \verb|Tensorflow Keras API| \\ \midrule
Datasets & \verb|Digits| \& \verb|MNIST| \\ \midrule
%Training Samples, Testing Samples &  PL: (\verb|100|, \verb|100|), QK: (\verb|500|, \verb|100|) \\ \midrule
Encoding & \verb|Angle Embedding| \\ \midrule
Optimizer & \verb|Adam| \\ \midrule
Epochs & \verb|50| \& \verb|200| \\ \midrule
Batch Size, Learning Rate & \verb|5|, \verb|0.01| \\ \midrule
\end{tabular}%
\end{adjustbox}
\end{table}

\subsection{Evaluation of our Cutting Methodology on the Digits Dataset}

\begin{figure*}[!ht]
  \centering
  \includegraphics[width=\linewidth]{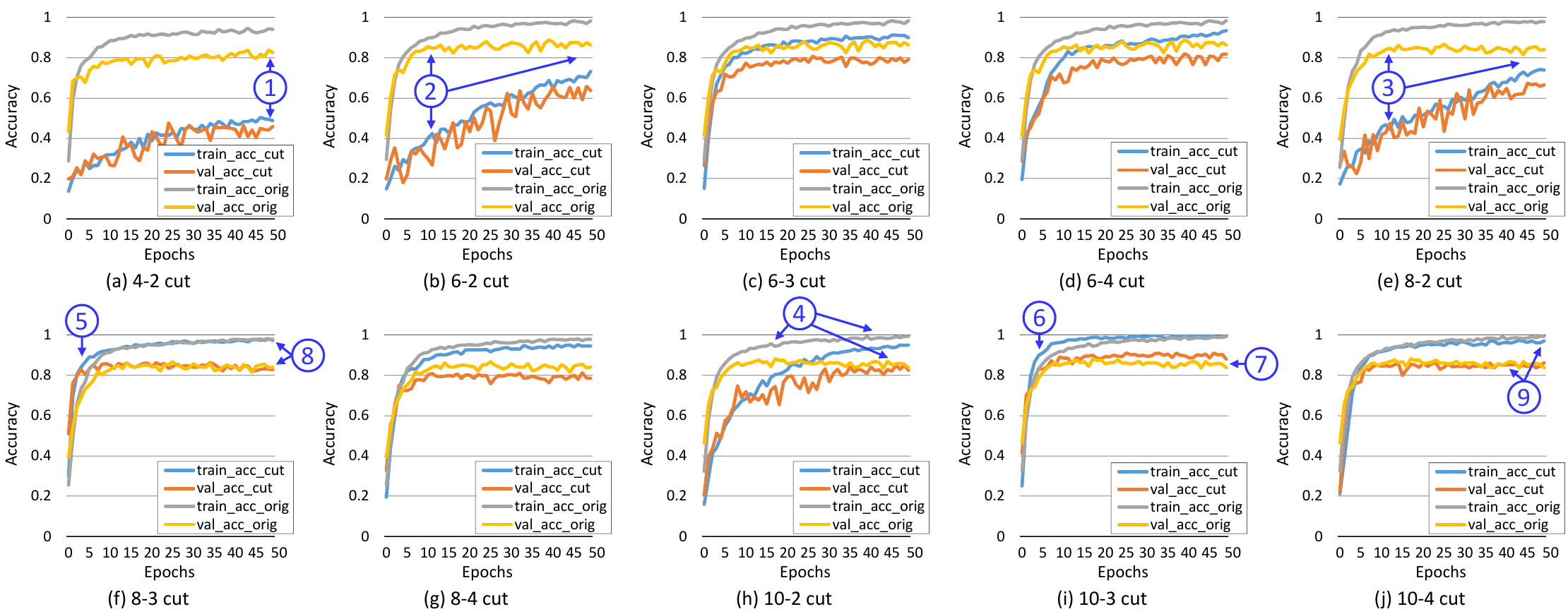}
  \caption{Training and validation accuracy for the original and cut circuits with different numbers of qubits on the Digits dataset. Each plot shows the results for a $n$-$m$ cut, with the $n$-qubit circuit implemented on an $m$-qubit device (using cutting) and compared with the original $n$-qubit circuit.}
  \label{fig:results_Digits}
  %\vspace{5pt}
\end{figure*}

The results of cutting the circuits for the Digits dataset are shown in \Cref{fig:results_Digits}. The results contain a set of $n$-$m$ cut experiments, comparing the training and validation accuracy of the original $n$-qubit circuit with the cut circuit implemented on an $m$-qubit device. For different values of $m$ and $n$, different observations can be derived. As indicated in labels~\rpoint{1},~\rpoint{2}, and~\rpoint{3}, for $4$-$2$, $6$-$2$, and $8$-$2$ cuts there is quite a large gap between the training accuracy of the original circuit and the training accuracy of the cut circuit. A similar large gap can be observed for the validation accuracy. The reason for this discrepancy can be attributed to the fact that when the circuits are implemented on $2$ qubits, the $2$-qubit subcircuits have low correlation and entanglement capabilities, which hinder the correct generalization of the problem.

However, for the $10$-$2$ cut circuit, the results are slightly different (see label~\rpoint{4}). While the training and validation accuracy gaps between the original circuit and the cut circuit are quite large after a few epochs, they shrink towards the end of the training epochs. This behavior suggested that the training process of the cut circuits implemented on a $2$-qubit device might require more training epochs to reach higher accuracy. Hence, in \Cref{subsec:results_more_epochs}, we analyze the behavior of these cut circuits when the HQNNs are trained for more epochs.

On the other hand, \textit{for devices that have more than $2$ qubits, the training and validation accuracy of the cut circuits are comparable with the original circuit}. As highlighted in labels~\rpoint{5} and~\rpoint{6} in \Cref{fig:results_Digits}, for the $8$-$3$ cut and $10$-$3$ cut circuits, during the first few epochs, the training and validation accuracy of the cut circuits are higher than the original circuits, meaning that the cut circuits converge faster than the original circuit. Moreover, the validation accuracy of the cut circuit for the $10$-$3$ cut even results slightly higher than the validation accuracy of the original circuits (see label~\rpoint{7}). This can be attributed to a better generalization when the cuts are deployed in the circuit. For other cases (see labels~\rpoint{8} and~\rpoint{9}), the training and validation accuracy curves for the original and cut circuits are overlapping, indicating that there is no significant functional difference between the two implementations.

\subsection{Evaluation of our Cutting Methodology on the MNIST Dataset}

\begin{figure*}[!ht]
  \centering
  \includegraphics[width=\linewidth]{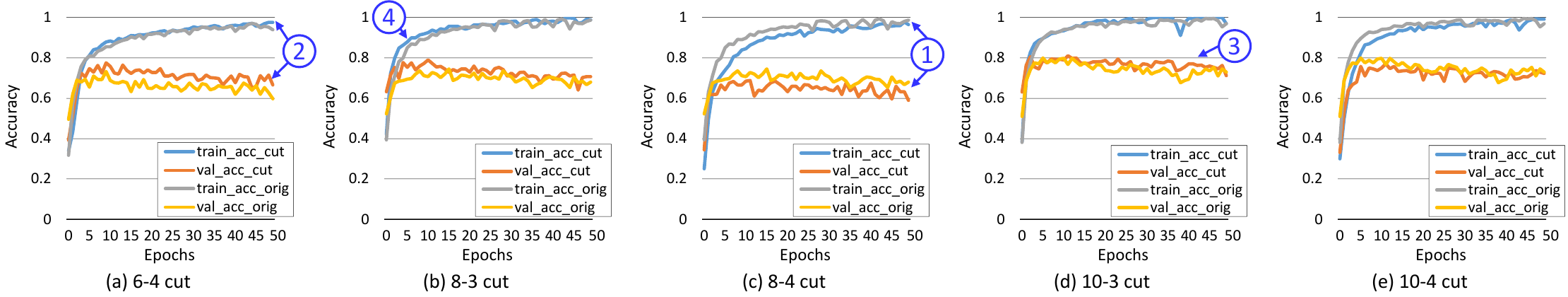}
  \caption{Training and validation accuracy for the original and cut circuits with different numbers of qubits on the MNIST dataset.}
  \label{fig:results_MNIST}
  %\vspace{5pt}
\end{figure*}

\Cref{fig:results_MNIST} shows the comparative results between the cut circuits and original circuits for the MNIST dataset. Overall, we can observe that the training and validation accuracy curves are similar for all the configurations. In some cases, for instance, for the $8$-$4$ cut (see label~\rpoint{1} in \Cref{fig:results_MNIST}), the training and validation accuracy of the cut circuit is slightly lower than that of the original circuit. In other cases, for the $6$-$4$ cut and $10$-$3$ cut circuits (see labels~\rpoint{2} and~\rpoint{3}), the training and validation accuracy of the cut circuit is slightly higher than that of the original circuit. Moreover, we can observe that for the first few epochs, the training and validation accuracies for the $8$-$3$ cut circuit are higher than for the original circuit (see label~\rpoint{4}).

Overall, the trends derived from the results conducted on both the Digits and MNIST datasets reveal good scalability and generalization for different benchmarks and workloads.

Note: we observe a certain level of overfitting in all the experiments, because the validation accuracy is significantly lower than the training accuracy. This phenomenon is well-known in neural networks and can be mitigated through techniques like dropout, regularization, and parameter adaptation~\cite{ChihChieh_2021TQC_ExpressibilityOverfittingQ, Kobayashi_2022QMI_EntanglingDropout, Scala_2023AQT_QNNDropout, Shinde_2024arxiv_QCNNParameterAdaptationOverfitting}. However, tackling it is beyond the scope of this work, since we just demonstrate that our quantum circuit cutting approach can be applied to HQNNs.

\subsection{Evaluation of the Impact of FLOPs Variations}

To quantitatively measure the computational overhead when implementing the cuts, we measured the FLOPs required for both the forward and backward passes, for different configurations of original and cut circuits. \Cref{tab:results_FLOPS} shows the results of our experiments on the Digits dataset. We can observe that, for all the configurations, cutting the circuits incurs an overhead in terms of FLOPs for the forward and backward passes. This overhead increases when reducing the number of qubits $n$ of the target device. This behavior is justified by the fact that the computations in the cut circuit require additional measurement and encoding operations of the intermediate results, to store and retrieve the partial states when executing the computations across multiple subcircuits. 

\begin{table*}[!ht]
\centering
\caption{Results of forward (FW), Backward (BW), and total (Tot) number of FLOPs for different versions of the original and cut circuits for the Digits dataset.}
\label{tab:results_FLOPS}
\begin{adjustbox}{max width=\linewidth}
\begin{tabular}{c|c|c|c|c|c|c|c|c|c|c|c|c|c|c}
Circuit & 4 & 4-2 & 6 & 6-2 & 6-3 & 6-4 & 8 & 8-2 & 8-3 & 8-4 & 10 & 10-2 & 10-3 & 10-4 \\ \toprule
FW FLOPs  & 48  & 96  & 72  & 192 & 132 & 108 & 96  & 320  & 208 & 160 & 120 & 480  & 300 & 240\\
BW FLOPs  & 230 & 306 & 404 & 554 & 494 & 478 & 602 & 866  & 754 & 716 & 852 & 1242 & 1062 & 1014\\
Tot FLOPs & 278 & 402 & 476 & 746 & 626 & 586 & 698 & 1186 & 962 & 876 & 972 & 1722 & 1362 & 1254
\end{tabular}
\end{adjustbox}
\end{table*}

% \subsection{Results: 4-Qubit Circuit on 2-Qubit Device - Digit dataset}

% \subsection{Results: 6-Qubit Circuit on 2-Qubit Device - Digit dataset}

% \subsection{Results: 6-Qubit Circuit on 3-Qubit Device - digit dataset}

% \subsection{Results: 6-Qubit Circuit on 4-Qubit Device - both MNIST \& digit dataset}

% \subsection{Results: 8-Qubit Circuit on 2-Qubit Device- both MNIST \& digit dataset}

% \subsection{Results: 8-Qubit Circuit on 3-Qubit Device- both MNIST \& digit dataset}

% \subsection{Results: 8-Qubit Circuit on 4-Qubit Device- both MNIST \& digit dataset}

% \subsection{Results: 10-Qubit Circuit on 2-Qubit Device- both MNIST \& digit dataset}

% \subsection{Results: 10-Qubit Circuit on 3-Qubit Device- both MNIST \& digit dataset}

% \subsection{Results: 10-Qubit Circuit on 4-Qubit Device- both MNIST \& digit dataset}

\subsection{Evaluation of 20-5 Cut \& 50-5 Cut on the MNIST Dataset}

\begin{figure}[!t]
  \centering
  \includegraphics[width=\linewidth]{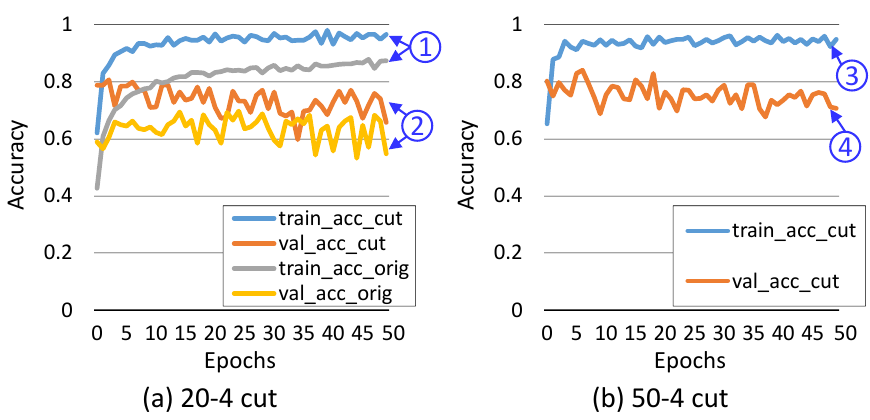}
  \caption{Training and validation accuracy results for 20 qubits, 20-4 cut, and 50-5 cut on the MNIST dataset.}
  \label{fig:Results_MNIST_20_50}
\end{figure}

\Cref{fig:Results_MNIST_20_50} presents the training and validation accuracy curves on the MNIST dataset for cut circuits derived from original circuits with 20 and 50 qubits\footnote{Only results from the cut circuit are reported for the 50-qubit case.}. All cut circuits are implemented using 4 qubits and Angle Embedding. Despite some fluctuations during training (see labels~\rpoint{2} and~\rpoint{4}), the curves demonstrate that the cutting technique remains effective even for larger-scale circuits, as accuracy converges to a high and stable level (see labels~\rpoint{1} and~\rpoint{3}). Notably, the 20-4 cut circuit shows a significant improvement in both accuracy and convergence speed compared to its uncut 20-qubit counterpart (see labels~\rpoint{1} and~\rpoint{2}), highlighting the advantages of circuit cutting.

\subsection{Ablation Study: Training Cut Circuits for More Epochs}
\label{subsec:results_more_epochs}

\begin{figure}[!t]
  \centering
  \includegraphics[width=\linewidth]{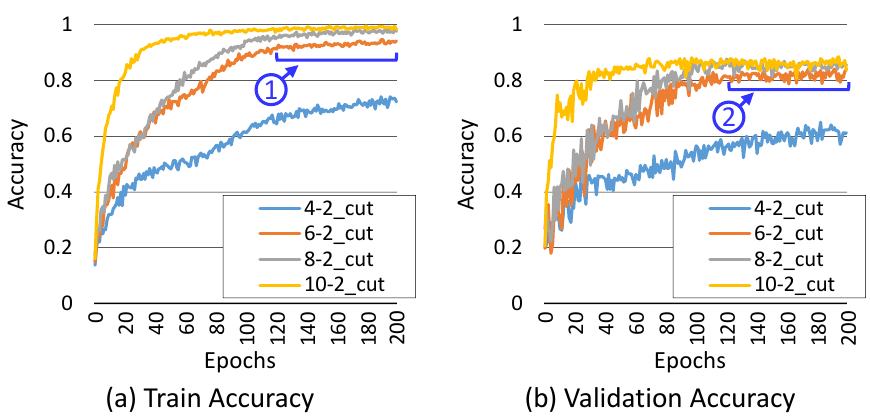}
  \caption{Training and validation accuracy of 4-2, 6-2, 8-2, and 10-2 cut circuits, trained for 200 epochs on the Digits dataset.}
  \label{fig:results_Digits_200epochs}
\end{figure}

Since the slope of the learning curves in some of the cut circuits is still high after 50 training epochs, we extended the training process to 200 epochs. \Cref{fig:results_Digits_200epochs} shows the training and validation accuracy curves for the $4$-$2$, $6$-$2$, $8$-$2$, and $10$-$2$ cut circuits. We can observe that the accuracy plateau is reached after 120 epochs for the $6$-$2$ and $8$-$2$ circuits (see labels~\rpoint{1} and~\rpoint{2} in \Cref{fig:results_Digits_200epochs}). For higher qubits of the original circuit (i.e., for the $10$-$2$ cut), the convergence appears after fewer epochs, while for lower qubits of the original circuits (i.e., for the $4$-$2$ cut), more training epochs are required to obtain the convergence.

\subsection{Ablation Study: Amplitude Embedding vs. Angle Embedding}
\label{subsec:results_amplitude}

\begin{figure}[!t]
  \centering
  \includegraphics[width=\linewidth]{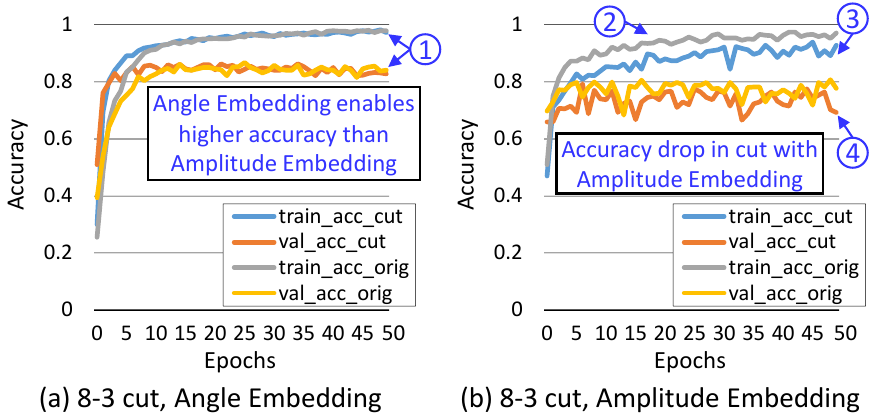}
  \caption{Comparison between Amplitude Embedding and Angle Embedding for 8-3 cuts on the Digits dataset.}
  \label{fig:Results_digits_amplitude}
\end{figure}

\Cref{fig:Results_digits_amplitude} presents a comparative analysis of encoding methods for mapping classical features into quantum states. The results indicate that Angle Embedding outperforms Amplitude Embedding, yielding higher learning curves. Specifically, the training and validation accuracy of the cut circuit with Angle Embedding closely match those of the original circuit (see label~\rpoint{1} in \Cref{fig:Results_digits_amplitude}). In contrast, Amplitude Embedding results in a noticeable accuracy decline even in the original circuit (see label~\rpoint{2}), with further degradation introduced by circuit cutting (see labels~\rpoint{3} and~\rpoint{4}). These findings underscore the superiority of Angle Embedding as a more effective encoding strategy for HQNN circuit cutting.

\subsection{Discussion}
\label{subsec:results_discussion}

The following key discussion points emerge from the experimental results:

\begin{itemize}[leftmargin=*]

\item \textbf{Feasibility of the Method:} The experiments demonstrated the feasibility of training Hybrid Quantum-Classical Neural Networks (HQNNs) on devices with limited qubit capacity. The experimental findings indicate that the cut circuits perform similarly to the original circuits, establishing quantum circuit cutting as an effective strategy for training and implementing HQNNs on devices with only a few qubits. An exception to this trend is observed in circuits cut to 2 qubits, where a slower convergence rate and slight losses in training and validation accuracy were noted compared to the original circuit. However, when employing 3 or more qubits, the cut circuits closely mirror the performance of the original circuits, sometimes even surpassing them in certain cases.

\item \textbf{Computational Overhead:} While there is a computational overhead associated with floating-point operations (FLOPS) in the cut circuits, the ability to execute the complete circuit on a reduced number of qubits significantly outweighs this cost. The reduced qubit requirement makes it possible to execute complex quantum circuits on hardware with limited resources, a critical advantage in the current NISQ era.

\item \textbf{Scalability:} This work serves as an initial proof-of-concept, demonstrating the practicality of training HQNNs on a limited number of qubits and proving the method's scalability when applied to cut circuits. The experimental setup allows for relatively fast and practical comparisons between cut and uncut circuits, revealing a trend of adaptability and scalability toward larger-scale quantum circuits. Although the present experiments focus on a small number of qubits to illustrate the concept, the results suggest that the methodology is scalable to more complex circuits. Future work will focus on extending this approach to larger circuits, further exploring its potential.

\end{itemize}

\section{Conclusion}

This work addresses the pressing challenges of executing large-scale quantum circuits on NISQ devices by introducing a novel quantum circuit cutting methodology for HQNNs. This method allows for the implementation and training of HQNNs on devices with limited qubit capacity while preserving the accuracy of the original circuits. The proposed approach not only facilitates the execution of large quantum circuits on constrained hardware but also enables the learning process for quantum parameters in all subcircuits, a critical component for the effective training of HQNNs.

Our experiments have demonstrated that the cut circuits perform comparably to the original circuits in terms of accuracy, with minimal computational overhead, especially when using three or more qubits. The scalability and adaptability of this method were also confirmed, showing potential for deployment on larger circuits, thus addressing a significant limitation of current quantum computing hardware.

This work enables the implementation and training of HQNNs on quantum devices that have limited qubits, in such a way that large-scale quantum circuits in HQNNs are split into multiple subcircuits with a lower qubit requirement. %Future work will focus on extending this methodology to even larger quantum circuits and exploring further optimizations to minimize computational overhead. 
The code and framework developed in this study will be made publicly available, ensuring reproducibility and fostering further advancements in the practical implementation of QML algorithms on NISQ devices.

\section*{Acknowledgment}
This work was supported in part by the NYUAD Center for Quantum and Topological Systems (CQTS), funded by Tamkeen under the NYUAD Research Institute grant CG008.

\begin{refsize}
\bibliographystyle{ieeetr}
\bibliography{main.bib}

\begin{thebibliography}{10}

\bibitem{Arute_2019Nature_QuantumSupremacy}
F.~Arute {\em et~al.}, ``Quantum supremacy using a programmable superconducting processor,'' {\em Nature}, 2019.

\bibitem{Araujo_2020SR_DivideAndConquer}
I.~F. Araujo {\em et~al.}, ``A divide-and-conquer algorithm for quantum state preparation,'' {\em Sci. Rep.}, 2020.

\bibitem{Huang_2022Science_QuantumAdvantageExperiments}
H.-Y. Huang {\em et~al.}, ``Quantum advantage in learning from experiments,'' {\em Science}, 2022.

\bibitem{Pokharel_2023PRL_QuantumSpeedup}
B.~Pokharel and D.~A. Lidar, ``Demonstration of algorithmic quantum speedup,'' {\em PRL}, 2023.

\bibitem{Kim_2023Nature_UtilityQuantumIBM}
Y.~Kim {\em et~al.}, ``Evidence for the utility of quantum computing before fault tolerance,'' {\em Nature}, 2023.

\bibitem{Tudisco_2024Access_AdvantagesQuantumFraudDetection}
A.~Tudisco {\em et~al.}, ``Evaluating the computational advantages of the variational quantum circuit model in financial fraud detection,'' {\em {IEEE} Access}, 2024.

\bibitem{kashif2022demonstrating}
M.~Kashif and S.~Al-Kuwari, ``Demonstrating quantum advantage in hybrid quantum neural networks for model capacity,'' in {\em ICRC}, 2022.

\bibitem{innan2025next}
N.~Innan {\em et~al.}, ``Next-generation quantum neural networks: Enhancing efficiency, security, and privacy,'' in {\em IOLTS}, 2025.

\bibitem{Preskill_2018arxiv_QC_NISQ_era}
J.~Preskill, ``Quantum computing in the {NISQ} era and beyond,'' {\em Quantum}, 2018.

\bibitem{kashif2021design}
M.~Kashif and S.~Al-Kuwari, ``Design space exploration of hybrid quantum--classical neural networks,'' {\em Electronics}, 2021.

\bibitem{Zaman_2024arxiv_QMLSurvey}
K.~Zaman {\em et~al.}, ``A survey on quantum machine learning: Current trends, challenges, opportunities, and the road ahead,'' {\em CoRR}, vol.~abs/2310.10315, 2024.

\bibitem{Chen_2022PRR_QCNN}
S.~Y.-C. Chen {\em et~al.}, ``Quantum convolutional neural networks for high energy physics data analysis,'' {\em Phys. Rev. Res.}, 2022.

\bibitem{Masum_2023ICMLA_HQNN_SentimentAnalysis}
A.~K.~M. Masum {\em et~al.}, ``Hybrid quantum-classical machine learning for sentiment analysis,'' in {\em ICMLA}, 2023.

\bibitem{LAbbate_2024TQE_QCTraining}
R.~L'Abbate {\em et~al.}, ``A quantum-classical collaborative training architecture based on quantum state fidelity,'' {\em IEEE TQE}, 2024.

\bibitem{Hu_2022ICCD_QGCNN_NISQ}
Z.~Hu {\em et~al.}, ``On the design of quantum graph convolutional neural network in the nisq-era and beyond,'' in {\em ICCD}, 2022.

\bibitem{Kashif_2024arxiv_InvestigatingNoiseHQNNs}
M.~Kashif {\em et~al.}, ``Investigating the effect of noise on the training performance of hybrid quantum neural networks,'' {\em CoRR}, vol.~abs/2402.08523, 2024.

\bibitem{kashif:2025_HQNET}
M.~Kashif and M.~Shafique, ``Hqnet: Harnessing quantum noise for effective training of quantum neural networks in nisq era,'' {\em arXiv:2402.08475}, 2025.

\bibitem{kashif2024nrqnn}
M.~Kashif and M.~Shafique, ``Nrqnn: The role of observable selection in noise-resilient quantum neural networks,'' in {\em World Congress in Computer Science, Computer Engineering \& Applied Computing}, 2024.

\bibitem{ahmed2025noisyhqnns}
T.~Ahmed, A.~Marchisio, M.~Kashif, and M.~Shafique, ``Noisy hqnns: A comprehensive analysis of noise robustness in hybrid quantum neural networks,'' in {\em IJCNN}, 2025.

\bibitem{ahmed2025quantum}
T.~Ahmed {\em et~al.}, ``Quantum neural networks: A comparative analysis and noise robustness evaluation,'' {\em arXiv preprint arXiv:2501.14412}, 2025.

\bibitem{McClean_2018arxiv_BarrenPlateaus}
J.~R. McClean {\em et~al.}, ``Barren plateaus in quantum neural network training landscapes,'' {\em CoRR}, vol.~abs/1803.11173, 2018.

\bibitem{kashif2023impact}
M.~Kashif and S.~Al-Kuwari, ``The impact of cost function globality and locality in hybrid quantum neural networks on nisq devices,'' {\em Machine Learning: Science and Technology}, 2023.

\bibitem{kashif2024resqnets}
M.~Kashif and S.~Al-Kuwari, ``Resqnets: a residual approach for mitigating barren plateaus in quantum neural networks,'' {\em EPJ Quantum Technology}, 2024.

\bibitem{kashif2024alleviating}
M.~Kashif {\em et~al.}, ``Alleviating barren plateaus in parameterized quantum machine learning circuits: Investigating advanced parameter initialization strategies,'' in {\em DATE}, 2024.

\bibitem{kashif2025deep}
M.~Kashif and M.~Shafique, ``Deep quanvolutional neural networks with enhanced trainability and gradient propagation,'' {\em Scientific Reports}, vol.~15, no.~1, p.~21764, 2025.

\bibitem{Peng_2020PRL_LargeQuantumCircuitsSmallQuantumCOmputer}
T.~Peng, A.~W. Harrow, M.~Ozols, and X.~Wu, ``Simulating large quantum circuits on a small quantum computer,'' {\em PRL}, 2020.

\bibitem{Tang_2021ASPLOS_CutQC}
W.~Tang {\em et~al.}, ``Cutqc: using small quantum computers for large quantum circuit evaluations,'' in {\em ASPLOS}, 2021.

\bibitem{Lowe_2023Quantum_FastQuantumCircuitCutting}
A.~Lowe {\em et~al.}, ``Fast quantum circuit cutting with randomized measurements,'' {\em Quantum}, 2023.

\bibitem{Piveteau_2024TIT_CircuitKnitting}
C.~Piveteau and D.~Sutter, ``Circuit knitting with classical communication,'' {\em IEEE TIT}, 2024.

\bibitem{Gentinetta_2024Quantum_CircuitKnitting}
G.~Gentinetta, F.~Metz, and G.~Carleo, ``Overhead-constrained circuit knitting for variational quantum dynamics,'' {\em Quantum}, 2024.

\bibitem{Bechtold_2023QST_EffectCircuitCuttingQAOANISQ}
M.~Bechtold {\em et~al.}, ``Investigating the effect of circuit cutting in qaoa for the maxcut problem on nisq devices,'' {\em QST}, 2023.

\bibitem{Brandhofer_2024TQE_PartitioningQuantumCircuitsGateCutsWireCuts}
S.~Brandhofer, I.~Polian, and K.~Krsulich, ``Optimal partitioning of quantum circuits using gate cuts and wire cuts,'' {\em IEEE TQE}, 2024.

\bibitem{Schmitt_2024arxiv_CuttingCircuitsTwoQubitUnitaries}
L.~Schmitt, C.~Piveteau, and D.~Sutter, ``Cutting circuits with multiple two-qubit unitaries,'' {\em CoRR}, vol.~abs/2312.11638, 2024.

\bibitem{Harrow_2024arxiv_OptimalQuantumCircuitCuts}
A.~W. Harrow and A.~Lowe, ``Optimal quantum circuit cuts with application to clustered hamiltonian simulation,'' {\em CoRR}, vol.~abs/2403.01018, 2024.

\bibitem{Bechtold_2024IPDPSW_CuttingWire}
M.~Bechtold, J.~Barzen, F.~Leymann, and A.~Mandl, ``Cutting a wire with non-maximally entangled states,'' in {\em IPDPSW}, 2024.

\bibitem{Chen_2024TQC_QuantumCircuitCuttingClassicalShadows}
D.~T.~S. Chen, Z.~H. Saleem, and M.~A. Perlin, ``Quantum circuit cutting for classical shadows,'' {\em ACM TQC}, 2024.

\bibitem{Kan_2024arxiv_ScalableCircuitCutting}
S.~Kan {\em et~al.}, ``Scalable circuit cutting and scheduling in a resource-constrained and distributed quantum system,'' {\em CoRR}, vol.~abs/2405.04514, 2024.

\bibitem{Sarker_2024IPDPSW_PerformanceWireCircuitCuttingEntanglements}
S.~Sarker {\em et~al.}, ``Quantifying performance of wire-based quantum circuit cutting with entanglements,'' in {\em IPDPSW}, 2024.

\bibitem{Pednault_2023arxiv_AlternativeWireCuttingAncillaQubits}
E.~Pednault, ``An alternative approach to optimal wire cutting without ancilla qubits,'' {\em CoRR}, vol.~abs/2303.08287, 2023.

\bibitem{Harada_2023arxiv_DoublyOptimalParallelWireCutting}
H.~Harada, K.~Wada, and N.~Yamamoto, ``Doubly optimal parallel wire cutting without ancilla qubits,'' {\em CoRR}, vol.~abs/2303.07340, 2023.

\bibitem{Bergholm_2018arxiv_PennyLane}
V.~Bergholm {\em et~al.}, ``Pennylane: Automatic differentiation of hybrid quantum-classical computations,'' {\em CoRR}, vol.~abs/1811.04968, 2018.

\bibitem{Alpaydin_1998UCI_DigitsDataset}
E.~Alpaydin and F.~Alimoglu, ``{Pen-Based Recognition of Handwritten Digits},'' {\em UCI MLR}, 1998.

\bibitem{Schuld_2018Springer_SupervisedQML}
M.~Schuld and F.~Petruccione, {\em Supervised Learning with Quantum Computers}.
\newblock Springer, 2018.

\bibitem{Rath_2023arxiv_QuantumDataEncoding}
M.~Rath and H.~Date, ``Quantum data encoding: A comparative analysis of classical-to-quantum mapping techniques and their impact on machine learning accuracy,'' {\em CoRR}, vol.~abs/2311.10375, 2023.

\bibitem{Arthur_2022QCE_HQNNs}
D.~Arthur and P.~Date, ``Hybrid quantum-classical neural networks,'' in {\em QCE}, 2022.

\bibitem{Bhowmik_2024arxiv_TransferLearningHQNN}
S.~Bhowmik and H.~Thapliyal, ``Transfer learning based hybrid quantum neural network model for surface anomaly detection,'' {\em CoRR}, vol.~abs/2409.00228, 2024.

\bibitem{Wang_2024ASPDAC_JustQ}
R.~Wang, F.~Baba-Yara, and F.~Chen, ``Justq: Automated deployment of fair and accurate quantum neural networks,'' in {\em ASP-DAC}, 2024.

\bibitem{Hafeez_2024AI_HQNN}
M.~A. Hafeez, A.~Munir, and H.~Ullah, ``H-qnn: A hybrid quantum–classical neural network for improved binary image classification,'' {\em AI}, 2024.

\bibitem{zaman2024comparative}
K.~Zaman {\em et~al.}, ``A comparative analysis of hybrid-quantum classical neural networks,'' in {\em World Congress in Computer Science, Computer Engineering \& Applied Computing}, 2024.

\bibitem{kashif2024computational}
M.~Kashif, A.~Marchisio, and M.~Shafique, ``Computational advantage in hybrid quantum neural networks: Myth or reality?,'' in {\em DAC}, 2025.

\bibitem{Senokosov_2024MLST_QMLImageClassification}
A.~Senokosov {\em et~al.}, ``Quantum machine learning for image classification,'' {\em Mach. Learn. Sci. Techn.}, 2024.

\bibitem{Zaman_2024arxiv_StudyingImpactQuantumHyperparameters}
K.~Zaman {\em et~al.}, ``Studying the impact of quantum-specific hyperparameters on hybrid quantum-classical neural networks,'' in {\em World Congress in Computer Science, Computer Engineering \& Applied Computing}, 2024.

\bibitem{Schillo_2024arxiv_QuantumCircuitLearningNISQ}
N.~Schillo and A.~Sturm, ``Quantum circuit learning on nisq hardware,'' {\em CoRR}, vol.~abs/2405.02069, 2024.

\bibitem{el2025designing}
W.~El~Maouaki {\em et~al.}, ``Designing robust quantum neural networks via optimized circuit metrics,'' {\em Advanced Quantum Technologies}, 2025.

\bibitem{Beer_2020Nature_TrainingDeepQNNs}
K.~Beer {\em et~al.}, ``Training deep quantum neural networks,'' {\em Nat. Commun.}, 2020.

\bibitem{Li_2023ICCD_VirtualDistillationCircuitCuttingMitigation}
P.~Li {\em et~al.}, ``Enhancing virtual distillation with circuit cutting for quantum error mitigation,'' in {\em ICCD}, 2023.

\bibitem{Majumdar_2022arxiv_ErrorMitigatedQuantumCircuitCutting}
R.~Majumdar and C.~J. Wood, ``Error mitigated quantum circuit cutting,'' {\em CoRR}, vol.~abs/2211.13431, 2022.

\bibitem{Basu_2024JSS_FragQC}
S.~Basu {\em et~al.}, ``Fragqc: An efficient quantum error reduction technique using quantum circuit fragmentation,'' {\em JSS}, 2024.

\bibitem{typaldos2024leveraging}
G.~Typaldos, W.~Tang, and J.~Szefer, ``Leveraging quantum circuit cutting for obfuscation and intellectual property protection,'' in {\em QCE}, 2024.

\bibitem{Marshall_2023Quantum_HighDimensionalQMLSmallQuantumComputers}
S.~C. Marshall, C.~Gyurik, and V.~Dunjko, ``High dimensional quantum machine learning with small quantum computers,'' {\em Quantum}, 2023.

\bibitem{Bilek_2024arxiv_UtilizingSmallQuantumComputersML}
S.~Bilek, ``Utilizing small quantum computers for machine learning and ground state energy approximation,'' {\em CoRR}, vol.~abs/2403.14406, 2024.

\bibitem{pira2022invitationdistributedquantumneural}
L.~Pira and C.~Ferrie, ``An invitation to distributed quantum neural networks,'' {\em CoRR}, vol.~abs/2211.07056, 2022.

\bibitem{Kawase_2024QMI_DistributedQNNs}
Y.~Kawase, ``Distributed quantum neural networks via partitioned features encoding,'' {\em QMI}, 2024.

\bibitem{Innan_2024arxiv_FedQNN}
N.~Innan {\em et~al.}, ``Fedqnn: Federated learning using quantum neural networks,'' in {\em IJCNN}, 2024.

\bibitem{Deng2012MNIST}
L.~Deng, ``The mnist database of handwritten digit images for machine learning research [best of the web],'' {\em IEEE SPM}, 2012.

\bibitem{ChihChieh_2021TQC_ExpressibilityOverfittingQ}
C.-C. Chen {\em et~al.}, ``On the expressibility and overfitting of quantum circuit learning,'' {\em ACM TQC}, 2021.

\bibitem{Kobayashi_2022QMI_EntanglingDropout}
M.~Kobayashi, K.~Nakaji, and N.~Yamamoto, ``Overfitting in quantum machine learning and entangling dropout,'' {\em QMI}, 2022.

\bibitem{Scala_2023AQT_QNNDropout}
F.~Scala, A.~Ceschini, M.~Panella, and D.~Gerace, ``{A General Approach to Dropout in Quantum Neural Networks},'' {\em AQT}, 2023.

\bibitem{Shinde_2024arxiv_QCNNParameterAdaptationOverfitting}
A.~R. Shinde, C.~Jain, and A.~Kalev, ``A post-training approach for mitigating overfitting in quantum convolutional neural networks,'' {\em CoRR}, vol.~abs/2309.01829, 2024.

\end{thebibliography}
\end{refsize}

\end{document}